\documentclass{article}
\usepackage{amssymb}
\def\abstract#1{\vskip 7mm 
        \begin{center}{\large Abstract}\par \smallskip
                \begin{minipage}[c]{12cm}
                        \small #1
                \end{minipage}
        \end{center}
}
\def\title#1{\begin{center}{\Large\bf #1}\end{center}}
\def\author#1{\vskip 5mm \begin{center}{#1}\end{center}}
\def\address#1{\begin{center}{\it #1}\end{center}}

\usepackage[dvips]{graphicx}
\usepackage{psfrag}

\psfrag{m}{$\frac{M}{\bar{M}}$}
\psfrag{rho}{$\log_{10}(\rho_{0(4)}/m^4)$}

\makeatletter
\@ifundefined{lesssim}{}{}
\@ifundefined{gtrsim}{}{}
\def\vereq#1#2{\lower3pt\vbox{\baselineskip1.5pt \lineskip1.5pt
\ialign{$\m@th#1\hfill##\hfil$\crcr#2\crcr\sim\crcr}}}
\makeatother

\begin{document}

\title{%
  Fermion Stars with an Extra Dimension
  \smallskip \\
  {}
}
\author{%
  Nahomi Kan\footnote{E-mail:b1834@sty.sv.cc.yamaguchi-u.ac.jp}
  and Kiyoshi Shiraishi\footnote{E-mail:shiraish@sci.yamaguchi-u.ac.jp}
}
\address{%
   Graduate School of Science and Engineering, Yamaguchi University, \\
  Yoshida,Yamaguchi-shi,Yamaguchi 753-8512, Japan
}

\abstract{
 Many efforts have been devoted to the studies of the phenomenology 
 in particle physics with extra dimensions.
 We propose the degenerate fermion star in the five dimensions, 
 and study 
 what effects caused by the geometry of extra dimensions should appear 
 in its structure.
 We note that Kaluza-Klein excited modes have effects 
 for the larger scale of extra dimensions 
 and examine the conditions on which different layers should be caused
 in the inside of the stars.
 We expound how the effects of the extra dimensions appears 
 on physical quantities.
}

\section{Introduction}
 Recently, physics dealing with higher dimensions has been noted 
 \cite{ref1,ref2}.
 Considering extra dimensions in GUT, 
 one points out four forces could be unified at low energy scale.
 And these are closely related with superstring theories,
 \footnote{The possibility of large extra dimensions was originally
 discussed by 
 Antoniadis\cite{anton}, with the relation to string theories. 
 The first string realization of low scale gravity models was given in \cite{aadd}.}
 supergravity, and the others.
 There are many works on extra dimensions.
 One of them is ``Neutron stars and extra dimensions'' 
 by Liddle, Moorhouse and Henriques \cite{ref3}.
 They considered the neutron stars in five dimensions 
 of which the fifth one is compactified into $S^{1}$, 
 and showed that the maximum mass of the stars decreases under the background.
 However there are some problems in their result.
 First, they did not consider Kaluza-Klein (K-K) excited modes 
 caused by the geometry of extra dimensions.
 Secondly, they put some hypotheses 
 both on the equation of state and on that of conservation, 
 so arbitrariness is left.
 Whereas, in this paper, 
 taking account of the K-K modes, 
 we propose the star in the five dimensions, 
 which is made of degenerate Dirac fermions, 
 and indicated that the excited modes have effects to the inside of star.
 After asking for the maximum mass and the radius 
 by use of numerical calculation,
 we will make it clear what the interior structure of star should be.

\section{Matter}

\subsection{(3+1) dimensions}
  In the four dimensional theory, 
  the thermodynamic potential of a fermion gas 
  with mass $m$ and half spin is

 \begin{equation}
 \Omega_{4}=-2\frac{1}{\beta}V
        \int\frac{d^{3}\textbf{p}}{(2\pi)^{3}}
        \left[\ln
        \left(1+e^{-\beta(\sqrt{\textbf{p}^{2}+m^{2}}-\mu)}
        \right)+(\mu \leftrightarrow -\mu)
        \right] ,
\end{equation}
 where we put subscript to emphasize 
 that it stands for a four dimensional quantity.
 We will restrict our system to degenerate fermion gas 
 and take the zero temperature limit.
 Then, the thermodynamic potential becomes

\begin{equation}
 \Omega_{4}(m) \equiv
       \left\{
       \begin{array}{@{\,}ll}
       -V\frac{m^{4}}{24\pi^{2}}
       \left[\frac{\mu}{m}\sqrt{\frac{\mu^{2}}{m^{2}}-1}
       \left(2\frac{\mu^{2}}{m^{2}}-5
       \right)+3\ln\left|\frac{\mu}{m}+
       \sqrt{\frac{\mu^{2}}{m^{2}}-1}
       \right| \right]  & (m<\mu)  \\
       0, & (m\geqq\mu)  
       \end{array}  .
       \right. 
 \label{eq:ome} 
\end{equation}
 Thermodynamical quantities follow from Eq.(\ref{eq:ome}).

\subsection{(4+1) dimensions}
  We will extend our argument into the five dimensional theory.
  Here we suppose that the fifith dimension is compactified into $S^{1}$  
  with a radius $b$, $b$ is not so large.
  We impose the periodic boundary condition on a wave function 
  in the fifth dimension:

\begin{equation}
 \psi(\chi)\sim{e^{ip_{5}\cdot\chi}} ~ ,
\end{equation}
\begin{equation}
 \psi(\chi+2\pi{b})\sim\psi(\chi)  ~ ,
\end{equation}
 with the fifth coordinate $\chi$ and momentum $p_{5}$ , then

\begin{equation}
 p_{5}=\frac{n}{b}  \ \ \ \mbox{($n$: integer)} ~ .
\end{equation}
 Therefore the relativistic energy involving the fifth dimension is

\begin{eqnarray}
 E_{5}&=&\sqrt{\textbf{p}^{2}+(\frac{n}{b})^{2}+m^{2}} \\
      &=&\sqrt{\textbf{p}^{2}+M^{2}} ~ , 
 \label{eq:e}
\end{eqnarray}
 where

\begin{equation}
 M^2\equiv\left(\frac{n}{b}\right)^{2}+m^{2} ~ ,
\end{equation}
($n$ is integer. Unless $n=0$, the K-K modes become effective.)
 Using Eq.(\ref{eq:e}),
 the thermodynamic potential of the fermion gas with mass $m$ and half spin 
 in the (4+1) dimensional space-time is

\begin{equation}
 \Omega_{5}=-2\frac{1}{\beta}V_{4}
        \int\frac{d^{4}\textbf{p}}{(2\pi)^{4}}
        \left[\ln
        \left(1+e^{-\beta(\sqrt{\textbf{p}^{2}+M^{2}}-\mu)}
        \right)+(\mu \leftrightarrow -\mu)
        \right] ~ ,
\end{equation}
 If we turn the integral over the fifth dimensional momentum 
 into the sum over quantum number $n$, that is :

\begin{equation}
 \int dp_{5} \to \frac{1}{b}\sum_{n} ~ ,
\end{equation}
 and in addition,

\begin{equation}
 V_{4}=2\pi{b}V  ~ ,
\end{equation}
 then the thermodynamic potential is

\begin{eqnarray}
 \Omega_{5}&=&-2\frac{1}{\beta}V\sum_{n}
        \int\frac{d^{3}\textbf{p}}{(2\pi)^{3}}
        \left[\ln
        \left(1+e^{-\beta(\sqrt{\textbf{p}^{2}
        +\frac{n^{2}}{b^{2}}+m^{2}}-\mu)}
        \right)+(\mu \leftrightarrow -\mu)
        \right] \nonumber \\
        &=&
        \sum_{n}\Omega_{4}
        \left(\sqrt{m^{2}+\frac{n^2}{b^2}}
        \right)=\sum_{n}\Omega_{4}(M) ~ .
\end{eqnarray}
 As well as the previous section, we will deal with the degenerate fermion gas 
 and low temperature near to zero.
 The thermodynamic potential, therefore,

\begin{equation}
 \Omega_{5}=
 \left\{
 \begin{array}{@{\,}ll}
 -Vb^{-4}f(\mu{b},mb), & (\mu > M)  \\
 0, & (\mu \leqq {M}) 
 \end{array} ~ ,
 \right.
 \label{eq:ome2}
\end{equation}
 where

\begin{equation}
 f\equiv\sum^{\infty}_{n=-\infty}\frac{(Mb)^4}{24\pi^2}
 \left\{\frac{\mu{b}}{Mb}\left(
 2\left(\frac{\mu{b}}{Mb}\right)^2-5\right)
 \sqrt{\left(\frac{\mu{b}}{Mb}\right)^2-1}+3\ln\left|
 \sqrt{\left(\frac{\mu{b}}{Mb}\right)^2-1}+\frac{\mu{b}}{Mb}
 \right|
 \right\} \ \ \   
 (\mu>M) ~ .
\label{eq:f}
\end{equation}
 From Eq.(\ref{eq:ome2}), thermodynamical quantities are as follows:

\begin{equation}
 P=-\frac{1}{2\pi{b}}\frac{\partial\Omega_{5}}{\partial{V}}
  =-\frac{1}{2\pi{b}}\frac{\Omega_{5}}{V}=
  \frac{1}{2\pi{b}}\frac{1}{b^{4}}f  ~ ,
\end{equation}

\begin{equation}
 P_{5}=-\frac{1}{2\pi{V}}\frac{\partial{\Omega_{5}}}{\partial{b}}
 =\frac{1}{2\pi{b}}\frac{1}{b^{4}}
 \left(x\frac{\partial{f}}{\partial{x}}+y\frac{\partial{f}}{\partial{y}}
 -4f \right) ~ ,
\end{equation}

\begin{equation}
 \rho=\frac{U}{V_4}=\frac{1}{2\pi{b}}\frac{1}{b^4}\left(
 x\frac{\partial{f}}{\partial{x}}-f \right) ~ ,
\end{equation}
 where $P_{5}$ is the pressure in the fifth dimension,
 and we put ~$x=\mu{b}$,~~$y=mb$.

\section{Space-Time}
 We take the line element to be of the form
 
\begin{equation}
 ds^2=e^{-\Phi}
 \left[-e^{-2\delta}\Delta{dt^2}+\frac{dr^2}{\Delta}+
 r^2(d\theta^{2}+\sin^{2}\theta{d\varphi^{2}})
 \right]
 +b_{0}^{2}e^{2\Phi}{d\chi}^2 ~ ,
\end{equation}
 where we regard $\Delta$, $\delta$, and $\Phi$ ($b=b_0e^{\Phi}$) 
 as functions depending only on $r$, 
 the distance from the origin.
 Next, the energy-momentum tensor is

\begin{equation}
 T^{\mu}_{\nu}=diag.(-\rho,P,P,P,P_{5}) ~ ,
\end{equation}
 in which we suppose isotropic pressure in the space of three dimensions,
 and represent the fifth dimensional pressure as $P_{5}$.
 The equation of conservation is

\begin{equation}
 \nabla_{\mu}T^{\mu{r}}=0 ~ .
\end{equation}
 From this, we find the chemical potential $\mu$ satisfies that

\begin{equation}
 \mu{b}=\frac{e^{\frac{3}{2}\Phi+\delta}}{\sqrt{\Delta}}\mu_{0}b_{0} ~ ,
\end{equation}
where $\mu$ depends on $r$ and 
 $\mu_{0}$ and $b_{0}$ are each the value of $\mu$
 and $b$ when $r$ is close to zero.  
  
\section{Equations}
 Now, we are just deriving Einstein equations.
 Before that, we will rewrite Eq.(\ref{eq:f}).
 After putting

\[
 x=\mu{b} ,~~~y=mb ,
\] 
 Eq.(\ref{eq:f}) reduces to
 
\begin{eqnarray}
 f(x,y)&=&{\sum^{\infty}_{n=-\infty}}'\frac{(y^{2}+n^{2})^2}{24\pi^{2}}
 \left[
 \frac{x}{\sqrt{y^{2}+n^{2}}}\sqrt{\frac{x^{2}}{y^{2}+n^{2}}-1}
 \left(2\frac{x^{2}}{y^{2}+n^{2}}-5\right) \right.   \nonumber \\
 & & \left.
 +3\ln\left|\frac{x}{\sqrt{y^{2}+n^{2}}}+\sqrt{\frac{x^{2}}{y^{2}+n^{2}}-1}
 \right|
 \right] \nonumber \\
 &=&{\sum^{\infty}_{n=-\infty}}'\tilde{f}(x,y)  \ \ \ \ \ \ \
 (\sqrt{x^{2}-y^{2}}>n^{2}) ~ ,
\end{eqnarray}
 where the sum over $n$ is done unless $n^{2}$ exceeds 
$\sqrt{x^{2}-y^{2}}$ .
 (To remark this, we put prime on a sum symbol.)
 In addition, we put

\begin{equation}
 Y=\sqrt{y^{2}+n^{2}} ~.
\end{equation}
 Using the leading formulae,
 Einstein equations are 

\begin{equation}
 \frac{{\tilde{M}_{\star}}'}{\tilde{r}^{2}}-\frac{3}{8}
 \tilde{\Delta}(\Phi')^{2}=
 4\pi\frac{1}{m^{4}b^{4}_{0}}e^{-6\Phi}{\sum}'
 \left(x\frac{\partial\tilde{f}}{\partial{x}}-\tilde{f} \right) ~ ,
\end{equation}

\begin{equation}
 \frac{1}{\tilde{r}}\delta'+
 \frac{3}{4}(\Phi')^{2}=
 -4\pi\frac{1}{m^{4}b^{4}_{0}}\frac{e^{-6\Phi}}{\tilde{\Delta}}{\sum}'
 \left(x\frac{\partial\tilde{f}}{\partial{x}} \right) ~ ,
\end{equation}

\begin{eqnarray}
 \Phi''+\left(\frac{\tilde{\Delta}'}{\tilde{\Delta}}-\delta'+
 \frac{2}{\tilde{r}} \right)\Phi'
 &=&-\frac{8\pi}{3}\frac{1}{m^{4}b^{4}_{0}}
 \frac{e^{-6\Phi}}{\tilde{\Delta}}{\sum}'
 \left(
 \frac{3Y^{2}-2y^{2}}{Y^2}x\frac{\partial\tilde{f}}{\partial{x}}
 \right.\nonumber  \\
 & & \left.-\frac{12Y^{2}-8y^{2}}{Y^2}\tilde{f}
 \right) ~ ,
\end{eqnarray}
where

\begin{equation}
 \tilde{M}_{\star}=\sqrt{G^{(4)}_{0}m^{2}}
 G^{(4)}_{0}mM_{\star} ~ ,
\end{equation}

\begin{equation}
 \tilde{r}=\sqrt{G^{(4)}_{0}m^{4}}r ~ ,
\end{equation}

\begin{equation}
 \tilde{\Delta}=1-\frac{2\tilde{M}_{\star}}{\tilde{r}} ~ .
\end{equation}
 In the above equations, we define a usual Newtonian constant 
$G_{0}^{(4)}$ as

\begin{equation}
 G_{0}^{(4)}\equiv\frac{G^{(5)}_{0}}{2\pi{b_{0}}} ~ ,
\end{equation}
 in which $G^{(5)}_{0}$ stands for the Newtonian constant 
 in the fifth dimension.
 And $M_{\star}$ is the mass of star depending on $r$.
 A prime means derivative with respect to $\tilde{r}$.
 We can solve these equations numerically.
 In the next section, we will show the result.          

\section{Results}
\begin{figure}[h]
\centering
\psfrag{m}{$\frac{M}{\bar{M}}$}
\psfrag{rho}{$\log_{10}(\rho_{0(4)}/m^4)$}
\includegraphics[width=13cm, height=7cm]{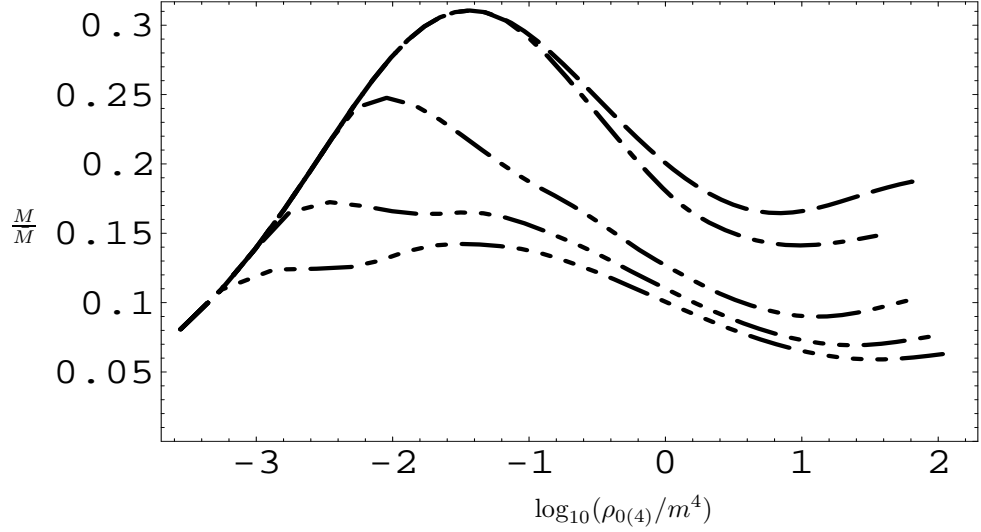}
\caption{
 Plots of the mass $M$ of the fermion stars 
 versus its central density $\rho_{0(4)}$ 
 for the various scales of the extra dimension. 
 The dashed line corresponds to $mb_0\approx 0$.
 The dot-dashed line corresponds to $mb_0=1$,
 the two dot-dashed line to $mb_0=2$,
 the three to $mb_0=3$,
 the four to $mb_0=4$.}
\label{fig:5-1}
\end{figure}

\begin{figure}[h]
\centering
\psfrag{m}{$\frac{M}{\bar{M}}$}
\psfrag{r}{$\frac{R}{\bar{R}}$}
\includegraphics[width=13cm, height=7cm]{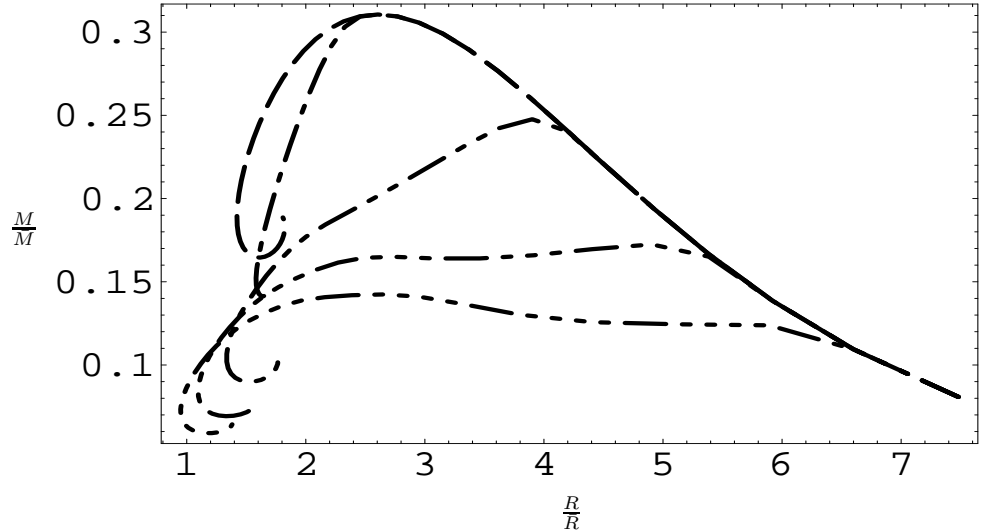}
\caption{
 Plots of the mass $M$ of the fermion stars
 versus its radius $R$ 
 for the various scales of the extra dimension.
 The correspondence between the dashed lines and the fifth dimension is the same as Fig.$\ref{fig:5-1}$.}
\label{fig:5-2}
\end{figure}

  We will exhibit the relationship between the mass and the central density 
 in fermion stars in Fig.$\ref{fig:5-1}$, 
 and the mass and the radius in Fig.$\ref{fig:5-2}$.
 It turns out that 
 the larger the extra dimension grows,
 the smaller the maximum mass becomes.
 Furthermore, it is remarkable 
 that two maximum points appeared when $mb_{0}=3,4$.
 
\newpage

\begin{figure}[p]
\centering
{\includegraphics[width=5cm, height=5cm]{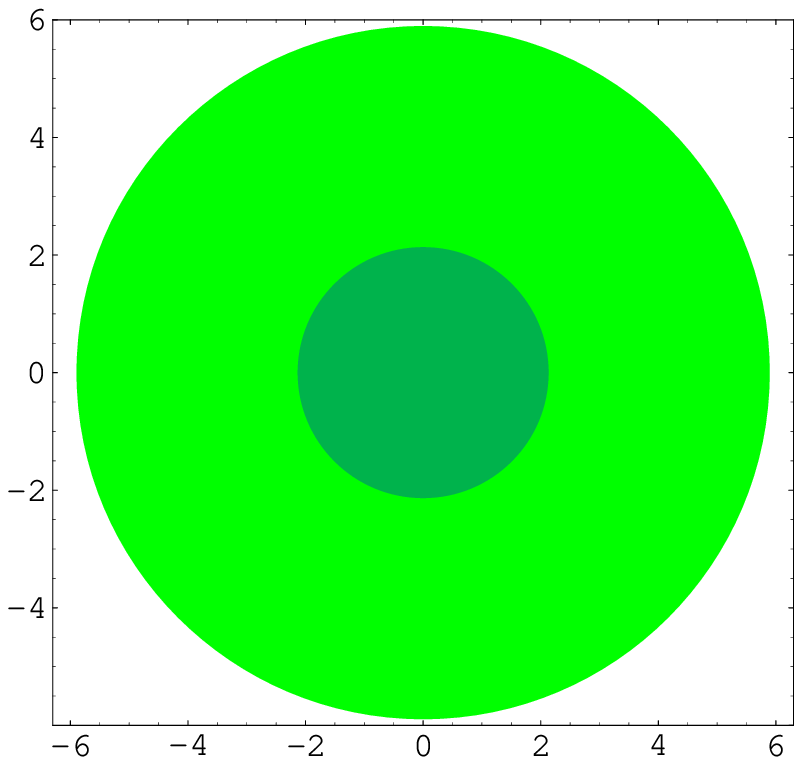}
 \hspace{0.3cm}\vspace{0.5cm}
 $mb_0=4$
 \includegraphics[width=5cm, height=5cm]{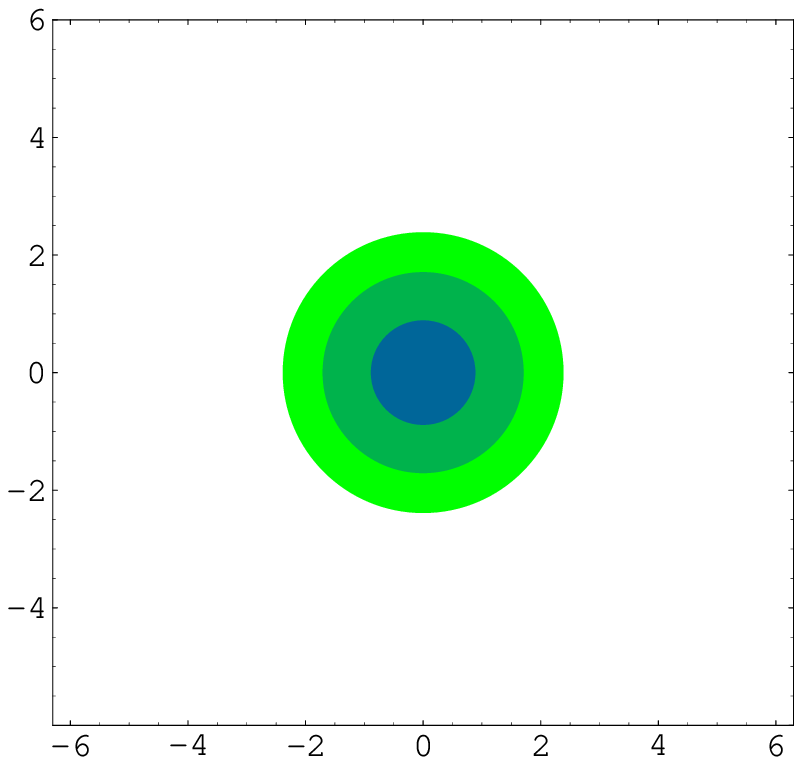}
 \hspace{1cm}\vspace{0.5cm}
 \includegraphics[width=5cm, height=5cm]{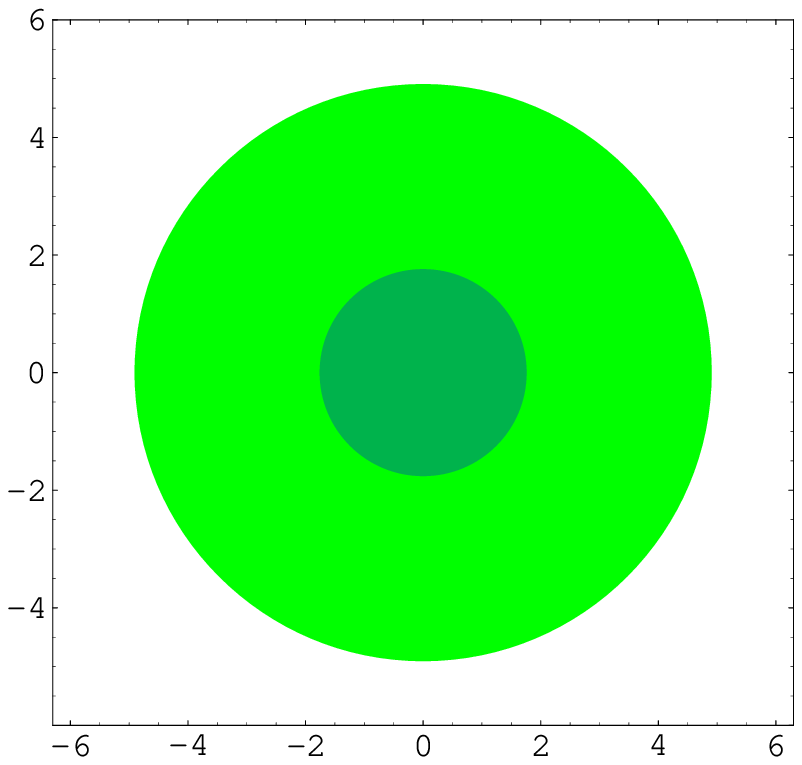}
 \hspace{0.3cm}\vspace{0.5cm}
 $mb_0=3$
 \includegraphics[width=5cm, height=5cm]{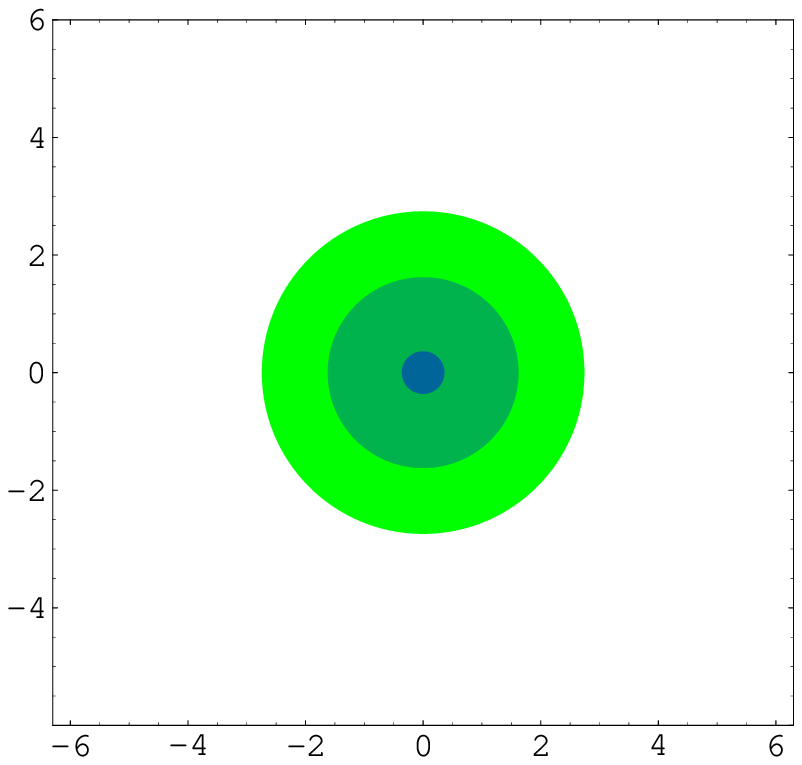}
 \hspace{12.5cm}\vspace{0.5cm}
 \hspace{1.3cm}
 \includegraphics[width=5cm, height=5cm]{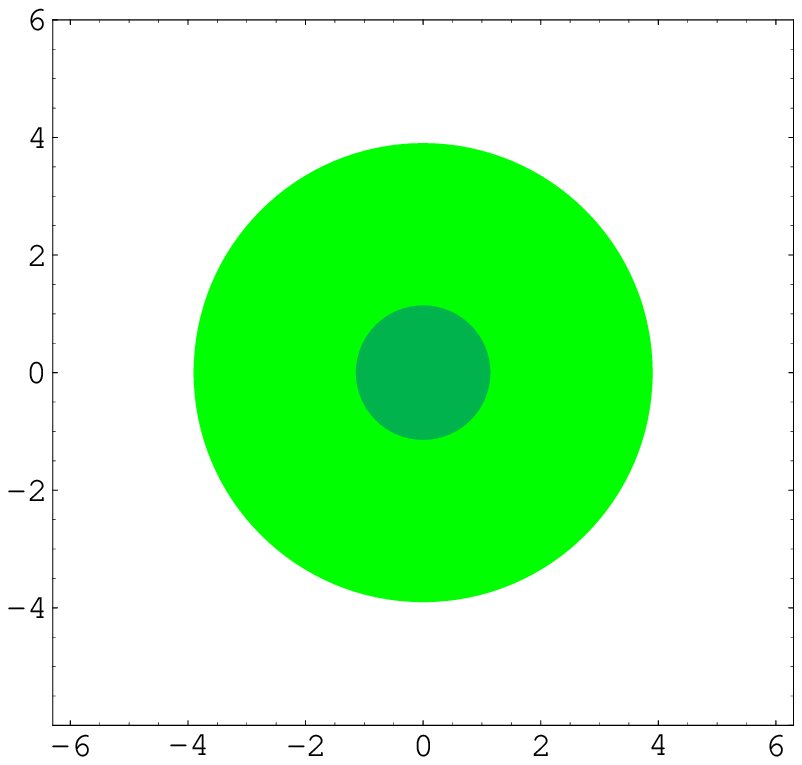}
 $mb_0=2$
 \hspace{12.5cm}\vspace{0.5cm}
 \hspace{1.3cm}
 \includegraphics[width=5cm, height=5cm]{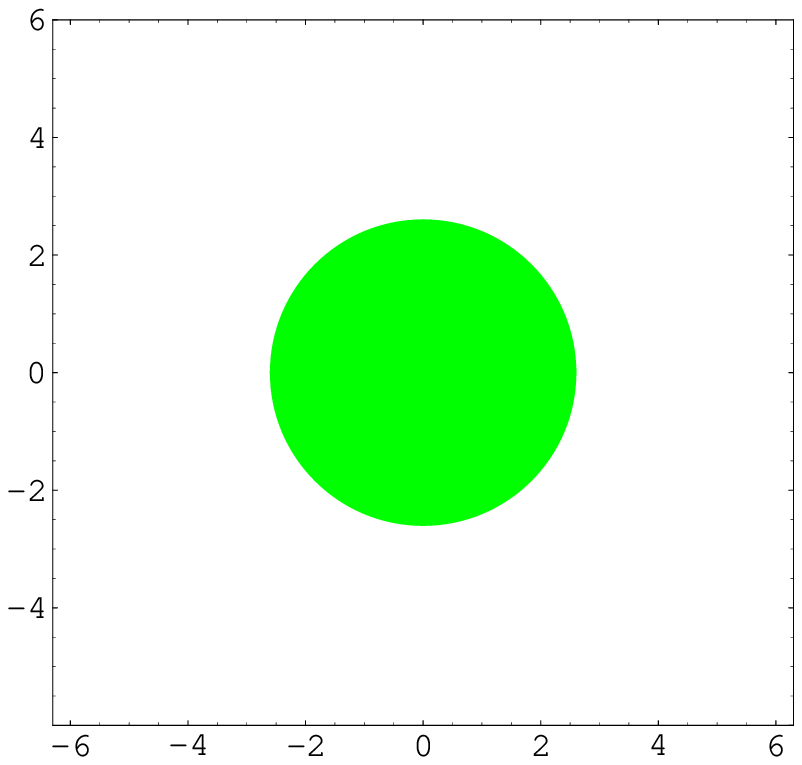}
 $mb_0=1$}
\caption{
 The evolutions of the fermion stars with the increase of the extra dimension.
 In the darkest circle, $n$=2 mode is most effective 
 and $n$=1 mode in the middle one.
 The pale circle stands for no excited mode.}
\label{fig:5-3}
\end{figure}

 Next, in Fig.$\ref{fig:5-3}$, we draw the interior structure of the stars 
 having the maximum mass, which is based on Fig.$\ref{fig:5-2}$.
 We can read off it from Fig.$\ref{fig:5-3}$ 
 that the excited modes have effects in the core of the stars 
 with increase in the extra dimension.
 We have two solutions in the case of $mb_{0}=3,4$.
 For $mb_{0}=3$, one of them is a larger star than that for $mb_{0}=2$
 and its central density is lower, which is shown in Fig.$\ref{fig:5-1}$.  
 While the other gets to be a smaller one and its central density is higher.
 In this solution, higher excited mode ($n$=2) is caused mainly 
 in the center of stars
 and lower mode ($n$=1) in the vast region including the core.
 Proceeding to $mb_{0}=4$, this inclination appears more remarkably.

\newpage 
\section{Conclusion}
 We have shown that considering K-K mode in the five dimensional theory, 
 as the extra dimension becomes larger,
 the maximum mass more and more decreases.
 And also we have obtained two series of solutions.
 One is that the fermion star is getting larger
 and its central density lower 
 with the increase of the extra dimension,
 that is, the larger and lighter star is formed.
 In these solutions, only $n$=1 mode is caused in the core of the stars.
 The other is that as the extra dimension grows,
 though the star once becomes larger and its central density lower 
 for $mb_0$=2,
 it is getting smaller and smaller and its central density higher 
 for the much larger dimension, 
 namely, the smaller and heavier star is created.
 In these stars, $n$=2,1 modes is caused, 
 the former is in the center of the stars
 and the latter is in the vast region including the core.
 As far as we examine, in this time,
 we have proved 
 that the structure of the fermion stars depends on 
 the scale of the fifth dimension, 
 that is, the excited modes have effects to the inside of stars.
 Taking the scale of the fifth dimension much larger,
 however,
 we need to analyze the stability of stars explicitly.
 And, in this paper, we have imposed the periodic boundary condition 
 on a wave function in the fifth dimension,
 while we can adopt the general one, that is: 
$\psi(\chi+2\pi{b})\sim{e^{i\varphi}}\psi(\chi)$.
 For the anti-periodic one,
$\psi(\chi+2\pi{b})\sim-\psi(\chi)$,
 the effective mass of the fermion $M_{(h)}$ is
$M_{(h)}=\sqrt{m^2+(\frac{n\pm\frac{1}{2}}{b})^2}$, 
 on which we are working,
 so, the details will be explained elsewhere.
 Meanwhile, we can apply our topics to cosmology 
 in which we will suppose the time dependence of the fifth dimension
 and think over how the stars should be created in the time-dependent process.
 Although we took the zero temperature limit in this paper, 
 we can treat the finite one too. 
 Furthermore we will try to extend our argument 
 into higher dimensional theories 
 (the six dimensions, the ten dimensions and the others),
 and consider what the star made of the bulk matter 
 in the brane world should be, 
 in which we are going to research as the next theme. 


\end{document}